\documentclass[sigconf]{acmart}
\AtBeginDocument{%
  }

\copyrightyear{2025}
\acmYear{2025}
\setcopyright{rightsretained}
\acmConference[CHI EA '25]{Extended Abstracts of the CHI Conference on Human Factors in Computing Systems}{April 26-May 1, 2025}{Yokohama, Japan}
\acmBooktitle{Extended Abstracts of the CHI Conference on Human Factors in Computing Systems (CHI EA '25), April 26-May 1, 2025, Yokohama, Japan}\acmDOI{10.1145/3706599.3720147}
\acmISBN{979-8-4007-1395-8/2025/04}

\begin{document}

\title{Fits like a Flex-Glove: Automatic Design of Personalized FPCB-Based Tactile Sensing Gloves}

\author{Devin Murphy}
\email{devinmur@mit.edu}
\orcid{0009-0000-9115-2146}
\affiliation{%
  \institution{MIT CSAIL}
  \city{Cambridge}
  \state{Massachusetts}
  \country{USA}
}

\author{Yichen Li}
\email{yichenl@mit.edu}
\orcid{0000-0002-5659-8748}
\affiliation{%
  \institution{MIT CSAIL}
  \city{Cambridge}
  \state{Massachusetts}
  \country{USA}}

\author{Crystal Owens}
\email{crystalo@mit.edu}
\orcid{0000-0002-2433-7025}
\affiliation{%
  \institution{MIT CSAIL}
  \city{Cambridge}
  \state{Massachusetts}
  \country{USA}}

\author{Layla Stanton}
\email{layla99@mit.edu}
\orcid{0009-0002-5425-7488}
\affiliation{%
  \institution{MIT}
  \city{Cambridge}
  \state{Massachusetts}
  \country{USA}}

\author{Young Joong Lee}
\email{youngyjl@mit.edu}
\orcid{0000-0002-2096-2261}
\affiliation{%
  \institution{MIT EECS}
  \city{Cambridge}
  \state{Massachusetts}
  \country{USA}}

\author{Paul Pu Liang}
\email{ppliang@mit.edu}
\orcid{0000-0001-7768-3610}
\affiliation{%
  \institution{MIT Media Lab and EECS}
  \city{Cambridge}
  \state{Massachusetts}
  \country{USA}}

\author{Yiyue Luo}
\email{yiyueluo@uw.edu}
\orcid{0009-0008-5127-0496}
\affiliation{%
  \institution{University of Washington ECE}
  \city{Seattle}
  \state{Washington}
  \country{USA}}

\author{Antonio Torralba}
\email{torralba@mit.edu}
\orcid{0000-0003-4915-0256}
\affiliation{%
  \institution{MIT CSAIL}
  \city{Cambridge}
  \state{Massachusetts}
  \country{USA}}

\author{Wojciech Matusik}
\email{wojciech@csail.mit.edu}
\orcid{0000-0003-0212-5643}
\affiliation{%
  \institution{MIT CSAIL}
  \city{Cambridge}
  \state{Massachusetts}
  \country{USA}}

\renewcommand{\shortauthors}{Murphy et al.}

\begin{abstract}
Resistive tactile sensing gloves have captured the interest of researchers spanning diverse domains, such as robotics, healthcare, and human-computer interaction. However, existing fabrication methods often require labor-intensive assembly or costly equipment, limiting accessibility. Leveraging flexible printed circuit board (FPCB) technology, we present an automated pipeline for generating resistive tactile sensing glove design files solely from a simple hand photo on legal-size paper, which can be readily supplied to commercial board houses for manufacturing. Our method enables cost-effective, accessible production at under \$130 per glove with sensor assembly times under 15 minutes. Sensor performance was characterized under varying pressure loads, and a preliminary user evaluation showcases four unique automatically manufactured designs, evaluated for their reliability and comfort.
\end{abstract}



\begin{teaserfigure}
  \includegraphics[width=\textwidth]{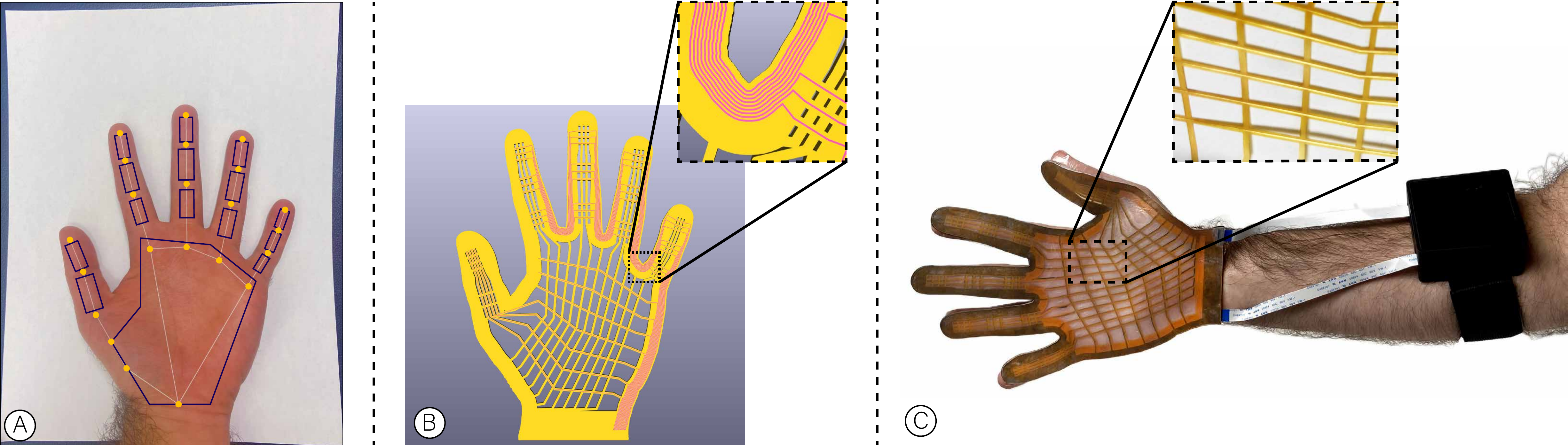}
  \caption{(A) A simple hand photo on legal-size paper is used to automatically generate manufacturing files (B) for personalized flexible printed circuit board tactile sensing gloves, ready to be ordered from a board house (C)}
  \Description{An overview of the pipeline with three panels side by side, labeled A through C. Panel A depicts a hand on legal-size paper, with hand landmarks overlaid as dots and polygonal sensing regions overlaid as purple strokes. Panel B depicts a 3d rendering of the FPCB design, with a close up on the electrodes routed along the finger regions. Panel C shows an outstretched hand wearing the FPCB tactile sensor.}
  \label{fig:teaser}
\end{teaserfigure}

\maketitle

\section{Introduction}
Human dexterous manipulation is a remarkable capability and has long captivated researchers across diverse fields~\cite{OkamuraDexterous2000, napier1993hands, rochat1989object}. Recently, tactile sensing gloves have emerged as powerful tools to extract meaningful insights from hand-environment interactions, enabling applications in robotic manipulation~\cite{Luo2024}, healthcare~\cite{TactileSensingFabrics2012}, and virtual reality~\cite{liu2023reconfigurabledataglovereconstructing}. Among the various sensing principles employed in these gloves, including capacitive, piezoelectric, and triboelectric mechanisms, resistive sensing stands out for its simplicity and scalability~\cite{Duan2022FLexiblePressureSensor}.

Resistive tactile sensing typically involves positioning a piezoresistive material between two orthogonally arranged electrode arrays. Sensing taxels are formed at the intersections of these arrays, where applied pressure induces measurable changes in the resistance of the piezoresistive layer. While several manufacturing methods have been proposed to create tactile sensing gloves following this principle ~\cite{Sundaram2019, Luo2021, Jiang2024}, most require either labor-intensive and error-prone manual assembly or access to expensive fabrication equipment.

In contrast, flexible printed circuit boards (FPCBs) have gained widespread adoption in wearable electronics due to their flexibility, durability, and compatibility with complex designs \cite{KuSkinLink2023, fpcbHaptics2017, Nan2022Review}. Features like instant quoting and integration with open-source electronic design automation (EDA) tools have further democratized access to FPCB technology, enabling rapid design iterations and streamlined production. By leveraging the growing accessibility and versatility of FPCBs, this work aims to make tactile sensing gloves more accessible to a broader research community.

To achieve this goal, we introduce a pipeline that automatically generates FPCB-based resistive sensing electrode arrays from a simple hand photograph on legal-size paper. Personalized to the unique contours of individual hands, these designs can be fabricated by commercial manufacturers—or "board houses"—at a cost of less than \$130 per glove. Assembly is straightforward, requiring less than 15 minutes and utilizing pressure-sensitive adhesive pre-applied to the components. To evaluate our approach, we characterized the sensors' response to varying pressure loads and conducted a preliminary user study with four personalized designs. Our results demonstrate sensitivity and repeatability comparable to commercial solutions while highlighting the potential benefits of personalization, such as improved signal consistency over repeated use and reduced obstructiveness. We conclude by discussing challenges related to FPCB durability under repeated bending and present an evaluation of recommendations from expert engineers to improve robustness in future designs.

\section{Related Work}
In this section, we describe related works in tactile sensing gloves and personalized sensor designs. 

\subsection{Tactile Sensing Gloves}
A tactile sensing glove captures touch sensations on the hand, contributing to studies in human behavior monitoring \cite{delpretoLiu2022actionSense}, robotics \cite{iroszhang2021}, and human-computer interaction \cite{liu2023reconfigurabledataglovereconstructing}. Recently, a few commercial tactile sensing gloves \cite{GripPressureSensorGlove, PressureProfileTactileGlove, NovelGloveForceMeasurement} are available, but they are relatively costly and have limited sizing options. \citet{Sundaram2019} introduced a high-resolution resistive tactile sensing glove by manually arranging orthogonal electrodes over a resistive layer, enabling the capture of a large and diverse dataset for human-object interactions. Other tactile sensing gloves have been fabricated using techniques like  manually integrated fluidics \cite{SpielbergFluidicPressure2020} and hand-sewn conductive yarns \cite{Jiang2024}. Compared to sensing mechanisms such as capacitive, piezoelectric, and triboelectric sensors \cite{Duan2022FLexiblePressureSensor}, resistive tactile sensing gloves offer notable advantages, including simple design, robust performance, and low power consumption. Further, to scale up the fabrication process, digital fabrication methods, including machine knitting \cite{Luo2021}, digital embroidery \cite{Luo2024}, and laser cutting \cite{Han2019Laser}, have been deployed to create resistive tactile sensing interfaces. However, current automated manufacturing processes face significant challenges, including the need for advanced equipment, difficulties in interfacing with readout electronics, and a reliance on one-size-fits-all designs. Flex-PCB technology has emerged as a promising solution to these limitations, with recent advancements in capacitive sensing \cite{CapacitivePressureSensorLi2024} and microelectromechanical systems \cite{straingauges2024}. In  this work, we present a flex-PCB approach for personalized resistive tactile sensing gloves, enabling mass manufacturing through established industries and making the technology accessible and customizable to a broader community.

% There are many ways to fabricate pressure sensors -> piezoresistive methods often chosen for scalability -> current automated manufacturing processes require advanced equipment, have difficulty interfacing with readout electronics, and assume a one-size fits all design. 

% Textile Integrated methods (STAG, Machine-knitted, Hand-sewn conductive threads)

\subsection{Personalized Sensor Design}
As stated by the "design for 1\%" principle \cite{norman2013design}, personalization is essential for crafting meaningful and effective designs. By embracing personalization, designers can deliver experiences that are more inclusive, adaptive, and user-friendly. \citet{Blom01092003} define personalization as adapting a system's functionality, interface, or content to enhance its relevance to an individual, improving ease of use, control, and perceived performance. Wearable sensor designs for applications such as electrical impedance tomography \cite{ZhuEITKit2021}, glucose monitoring \cite{LakhdirGlucoMaker2024}, and other biomedical uses \cite{Ventola2014Medical3dPrinting} benefit significantly from personalization, improving both user experience and sensor performance \cite{JarusriboonchaiCustom2019}. However, an automated interface tailored to tactile sensing gloves has yet to be developed, despite the variability in hand shapes from person to person. 
% \yy{TODO: check this part, not sure whether we wanna talk about diverse wire placement algorithms?}
% Designing resistive-matrix based tactile sensors to fit convex geometries currently requires considerable manual efforts to lay out electrode traces. While recent works have made progress in automatic matrix electrode placement for multi-touch capacitive sensing on general 3d surfaces. 

% Custom senso\citet{ZhuEITKit2021} presents custom electrode placement for Electrical Impedance Tomography sensing. \citet{LakhdirGlucoMaker2024} presents custom glucose monitors. 

% Theory of personalization, benefits of applying this to wearable sensor design -> Past works that have created interfaces for personalized design, or automated personalized design -> Such an automated interface has yet to be designed for tactile sensing gloves, despite the unique shape of the hand from person to person. 

% \input{03_method}
\section{Automated Design and Manufacturing Pipeline}

We present an algorithm that automates the design of personalized FPCB-based resistive tactile sensors, optimized for seamless integration with sensor readout electronics and fabrication by standard PCB manufacturers. The algorithm generates all necessary design components, including electrode traces, a coverlay mask, and an adhesive region. These sensors feature orthogonally arranged electrodes on a pair of top and bottom FPCBs, enabling straightforward assembly through the sandwiching of the two FPCBs, a middle piezoresistive layer, and silicone encapsulation layers on both sides.

\begin{figure*}
    \centering
    \includegraphics[width=1\linewidth]{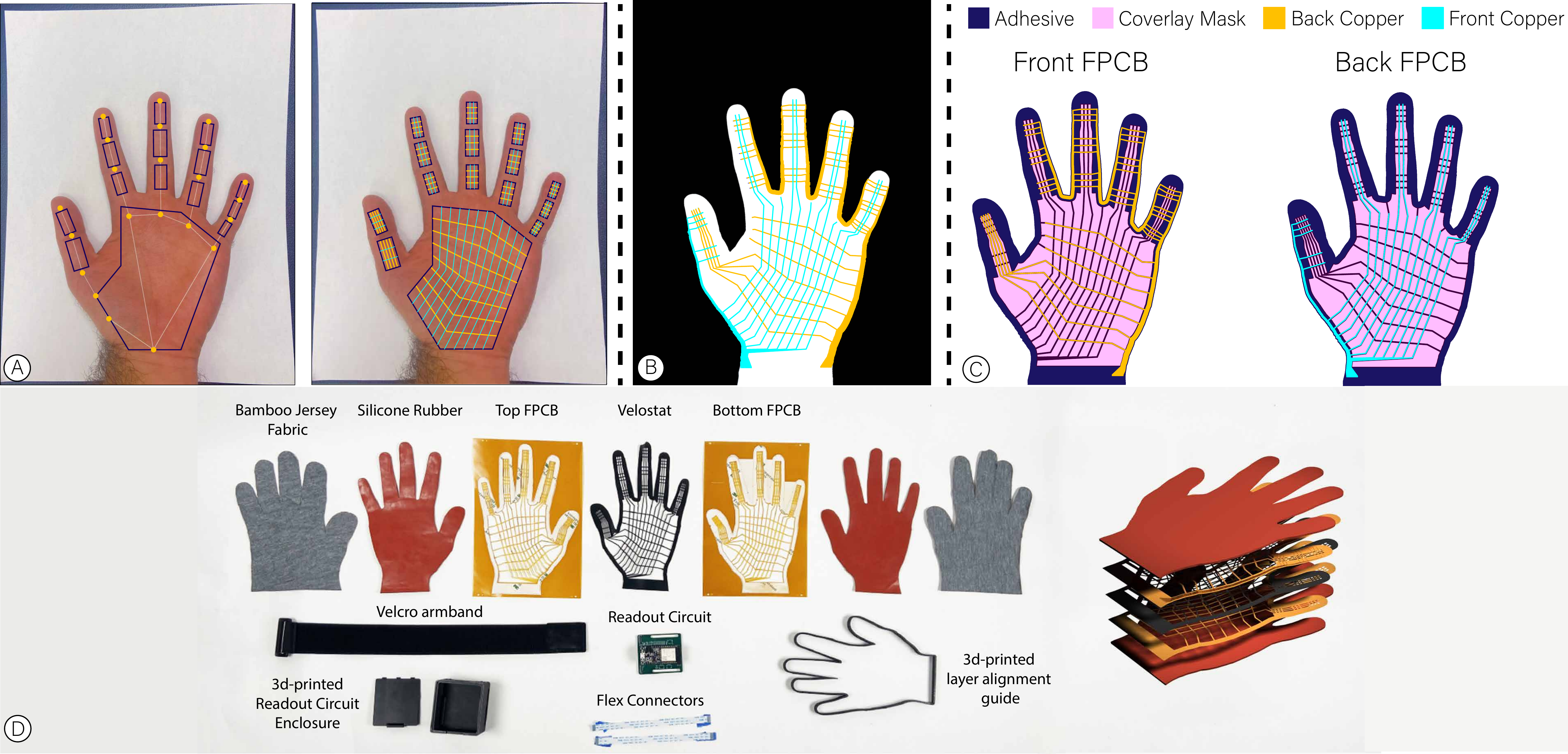}
    \caption{Automated FPCB design and sensor assembly. (A) Sensing regions are detected and orthogonally arranged electrodes are generated within them. (B) The binary mask of the hand is used to extract its contour, which serves to connect electrodes across neighboring regions. (C) Supporting layers of the PCB are generated and used to create Gerber files which can be sent directly to PCB manufacturers. (D) Sensor assembly using materials totaling under \$130 dollars by sandwiching 5 layers together with pressure-sensitive adhesive.}
    \label{fig:design-alg}
    \Description{The automated design and manufacturing pipeline is illustrated across four panels labeled A through D. Panel A shows two images of a hand on legal-size paper side by side. The image on the left depicts hand landmarks as dots and calculated polygonal sensing regions overlaid on top. The image on the right depicts the generated horizontal and vertical electrodes within the sensing region. Panel B displays the hand as a binary mask, with the fully connected horizontal and vertical electrodes overlaid. Panel C shows the final generated design, which is a Front and Back FPCB with the horizontal and vertical electrodes, respectively, and coverlay mask and adhesive layers. Panel D shows all materials for manufacturing laid out, including bamboo jersey fabric, silicone rubber, velostat, the FPCBs, a readout circuit, flex connectors, a 3d printed readout circuit enclosure, a 3d-printed layer alignment guide, and a velcro armband.}
\end{figure*}

\subsection{FPCB Design Algorithm}

Our algorithm takes as input an image of the user hand laid on a legal size paper and outputs the FPCB sensor design components that can be readily fabricated by manufacturers (Fig.~\ref{fig:design-alg}). We describe our algorithm as three steps: automatic hand landmark detection for sensing region electrode placements, binary hand mask and connecting trace generation, and supporting layer generation.

\paragraph{Hand Landmark and Electrode Placement} Our objective is to arrange orthogonally positioned electrodes within regions of interest on the hand, specifically each digital pad and the palm. Starting with an image of a hand placed on a legal-size sheet of paper (8.5" x 11"), we first calculate the pixel-to-inch conversion ratio to scale the design file. We then use the open-source MediaPipe library~\cite{zhang2020mediapipehandsondevicerealtime} to detect 21 hand landmarks, which form the 2D hand skeleton, and automatically generate an approximate layout for the 15 regions of interest (three for each finger, two for the thumb, and one for the palm). To define the finger regions, we compute the vector direction along each skeleton segment and generate a rectanglular sensing region along this direction, with the width set to half the segment's length. The length of the rectangles above the palm is determined using standard ratios from the metacarpophalangeal (MCP) joints to the proximal interphalangeal (PIP) joints for each finger~\cite{handratios2021}. The palm region is represented as a 7-point polygon, defined by projecting the hand landmarks in the palm area along various parts of the skeleton. Within each sensing region, we create orthogonally positioned horizontal and vertical traces by evenly sampling along the edges, generating unique sensing taxels at the intersections of these traces. For our initial design, we pre-define the resolution for each region (3x2 for each finger pad, 4x3 for the top of the thumb, 4x2 for the middle of the thumb, and 7x11 for the palm), totaling 167 unique pressure-sensing nodes. 

\paragraph{Binary Mask and Connecting Trace Generation} The generated electrodes must connect to a readout circuit, with all horizontal copper traces routed to a single connector and all vertical copper traces similarly routed.  We begin by connecting neighboring traces that can be directly linked without crossing the hand's contour. This includes connecting vertical traces along the finger regions to one another and to the palm, as well as connecting the four horizontal traces in the thumb regions to each other and to the bottom four horizontal traces of the palm. For the remaining connections, we route the traces along the hand's contour. To do this, we first generate a binary mask to segment the hand from the paper background using HSV thresholding, then automatically generate dense line segments representing the hand's contour from the mask. We set the trace thickness to 0.1 mm and apply slight inward extrusions of 0.2 mm to create 16 connecting traces for both the horizontal and vertical electrodes, ensuring no trace clearance requirements are violated. By connecting the bottom right-most electrodes to the innermost connecting trace and the top electrodes to the outermost trace, we link any unconnected electrodes in the sensing regions to the nearest points on the corresponding contour (Fig. \ref{fig:design-alg}B).

\paragraph{Supporting Layers Generation} The FPCB requires additional layers for functionality: the coverlay mask, adhesive, and edge cut layers (Fig. \ref{fig:design-alg}C). Coverlay is a polyimide film normally laminated over copper traces to provide flexibility and electrical insulation. However, resistive sensing requires direct electrode contact with the piezoresistive layer, so we define a coverlay mask as the convex hull of the sensing regions on the fingers with an inward extrusion of the hand contour. Adhesive layers, securing the piezoresistive layer, occupy the area between the hand's outer contour and the coverlay mask. The edge cut layer, marking where the board will be laser-cut, includes both the hand's outer contour and inner cuts to enhance flexibility and dexterity~\cite{permeabilityTeng2024}. These inner cuts are created by outwardly extruding copper traces by 0.5 mm (to stay within most manufacturer’s edge clearance requirements ~\cite{jlcpcb_flex_pcb}) and subtracting the union of horizontal and vertical trace polygons from the coverlay mask.

The horizontal and vertical traces along with their supporting layers are exported into two different SVG files representing the front FPCB and back FPCB of the sensor, and converted into KiCad PCB footprints \cite{kicad_footprints} using the open-source "gerbolyze" tool \cite{gerbolyze}. From KiCad, the Gerber files, component placement, and bill-of-materials for each PCB can be exported and sent to a board house for manufacturing.

\subsection{Sensor Assembly}

\begin{table}[h]
    \centering
    \begin{tabular}{l r}
        \toprule
        \textbf{Item} & \textbf{Cost (USD)} \\
        \midrule
        FPCBs & 24.93 \\
        Velostat & 2.50 \\
        Silicone rubber & 29.20 \\
        Fabric & 4.62 \\
        Readout Circuit & 64.66 \\
        Flex cables & 0.79 \\
        \midrule
        \textbf{Total} & \textbf{126.70} \\
        \bottomrule
    \end{tabular}
    \caption{Detailed cost breakdown for one glove}
    \label{tab:cost_breakdown}
\end{table}

\label{sec:manufacture}
The hand contour from the generated SVG file is used to laser cut two mirrored layers of silicone rubber (1/32" thickness, 10A durometer) with adhesive backing. The perimeter and internal cutouts are then used to laser cut a single layer of piezoresistive Velostat\textregistered. We created an offset 3D-printed model of the hand outline from the SVG file, which served as a guide for accurately aligning and assembling the five layers that make up each glove. Assembly requires only the bonding of these layers using pressure-sensitive adhesive already deposited on the FPCB and silicone layers by manufacturers. For our user evaluation (Sec. \ref{sec:user-eval}), we also used each hand's contour to sew a classic fabric glove. This process involved laser cutting three layers of fabric with seam allowances based on the hand's outline: two layers surrounding the sensor assembly and one forming the back of the glove. The three layers are sewn together, and the sensor is then inserted into place. The horizontal and vertical electrodes of the sensor connect to a readout circuit via 16 pin 0.5mm pitch flat flexible cables (FFCs). We 3d-printed an enclosure for the readout circuit for ease-of-use during evaluations (Fig. \ref{fig:design-alg}D). Table \ref{tab:cost_breakdown} shows a detailed cost breakdown of the necessary components for manufacturing, totaling under \$130 USD per glove.
\begin{figure}
    \centering
    \includegraphics[width=\linewidth]{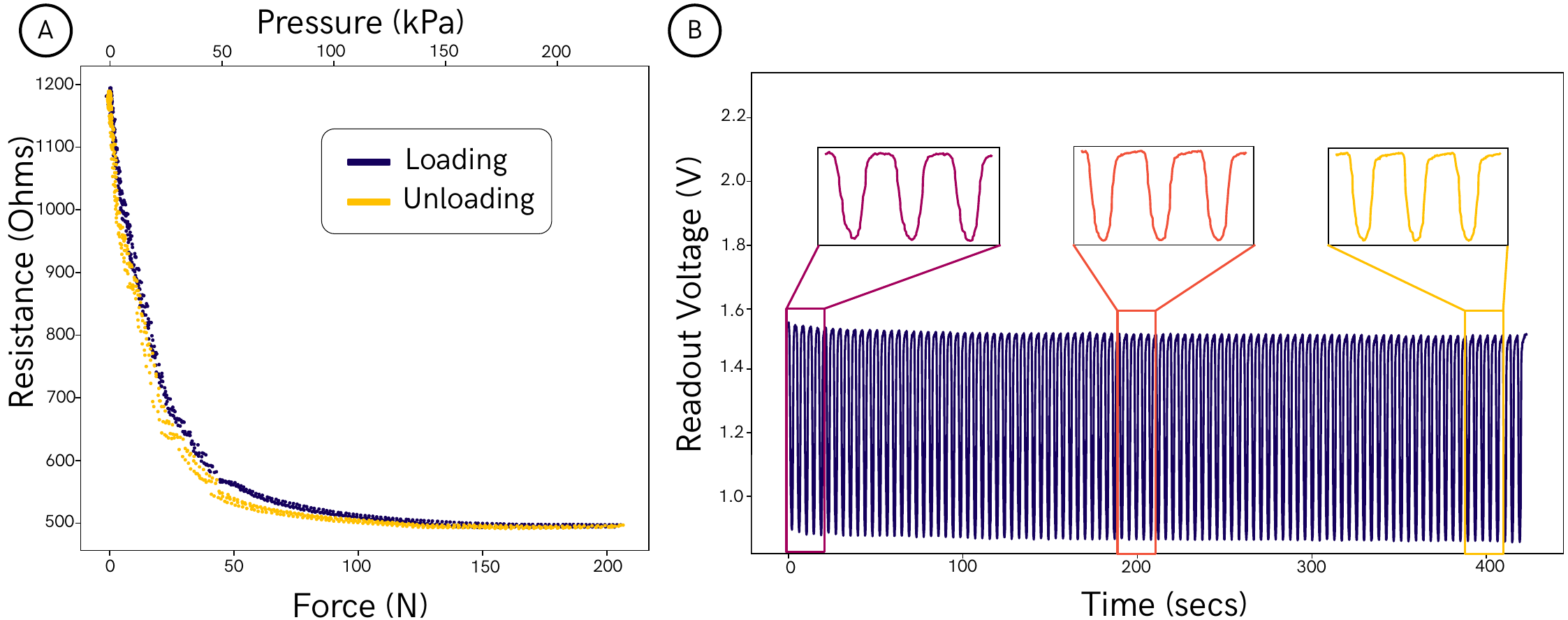}
    \caption{Sensor Characterizations: (A) Resistance variation during three loading and unloading cycles. (B) Consistent performance is captured by the readout circuit under a load of 12 $N/\text{cm}^2$ for 100 cycles.}
    \label{fig:characterization}
    \Description{Panel A: Line Graph showing resistance in ohms from 500 to 1200 on the Y axis versus force in newtons from 0 to 200 on the bottom X axis and pressure from 0 to 200 kPA on the top X axis. Two scatter plots show resistance logarithmically decreasing as applied force increases, for both loading and unloading cycles. A small bias between the two curves indicates some hysteresis. Panel B: Line Graph showing readout voltage from 0 to 2.2 volts on the Y axis with time from 0 to 400 seconds on the X axis. The readout voltage changes periodically and consistently across 100 cycles. }
\end{figure}

\section{Sensor Characterization}

We conducted two experiments to characterize the response of the FPCB sensor. First, a mechanical tester (Shimadzu AGX-V2) applied adjustable normal forces over a contact area of \( 9 \, \text{cm}^2 \). Using a digital multimeter (DMM), we recorded changes in resistance at a single node in the palm region across three loading and unloading cycles (Fig.~\ref{fig:characterization}A). The results demonstrated the expected inverse relationship between applied pressure and resistance, with two distinct linear regions and sensitivity up to 175~kPa, matching the performance of commercial solutions \cite{PressureProfileTactileGlove}. In the second experiment, a load of \( 12 \, \text{N/cm}^2 \) was applied repeatedly for 100 cycles over the same contact area. Voltage readings were acquired from an ESP32 microcontroller via its analog-to-digital converter (ADC), utilizing the open-source WiReSens Toolkit~\cite{murphy2024wiresenstoolkit}. The measurements were taken using a readout circuit based on a standard zero-potential architecture~\cite{DALESSIO199971}. The sensor exhibited stable performance across all cycles, demonstrating repeatability under extended use (Fig.~\ref{fig:characterization}B).

\section{Preliminary User Evaluation}

\begin{figure*}
    \centering
    \includegraphics[width=1\linewidth]{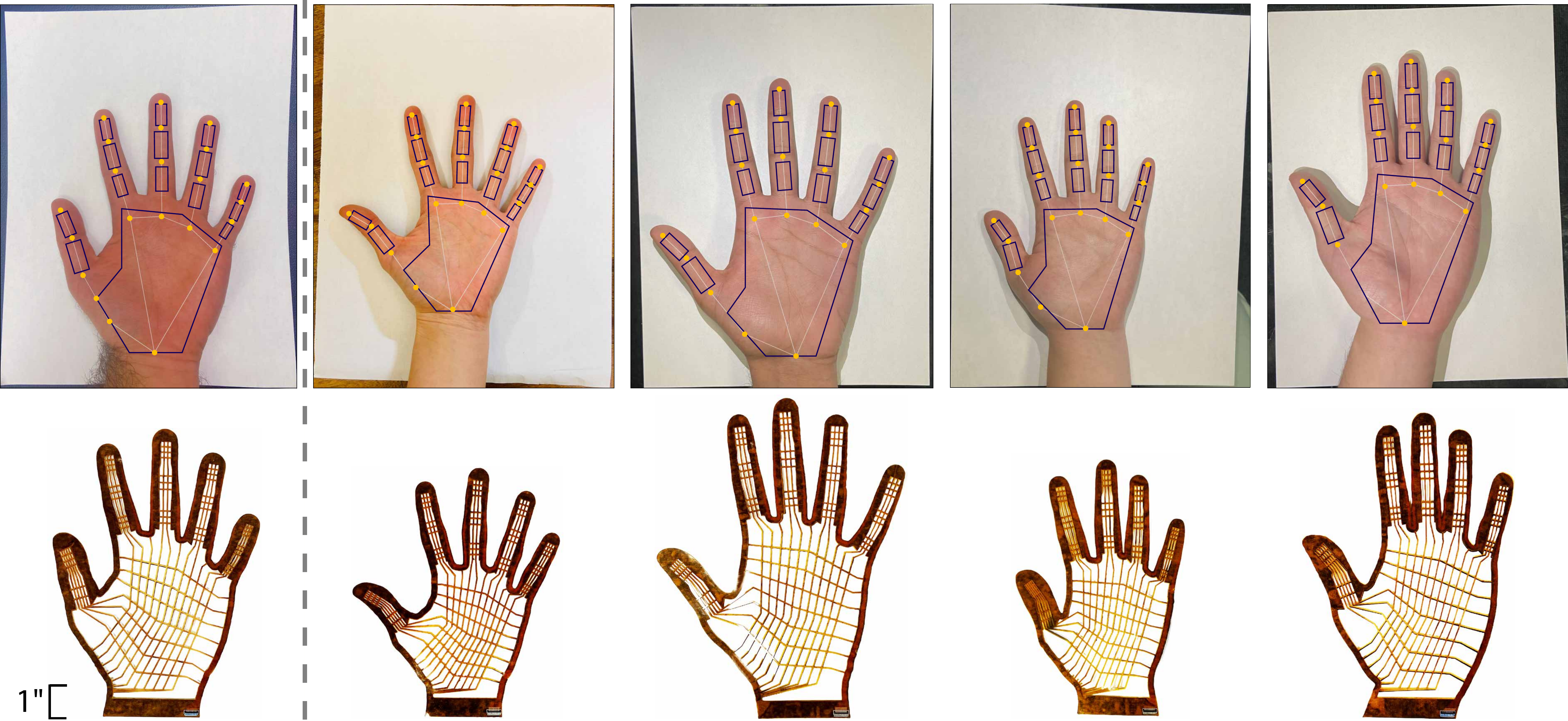}
    \caption{Automatically designed and manufactured FPCB tactile sensors for five individuals. Left: The sensor with median perimeter is used as the basis for a generic glove.}
    \label{fig:glove-designs}
    \Description{Manufactured FPCB designs in two rows. Top row: each of five hands on legal-size paper, with automatically determined sensing regions. Bottom row: each of the five automatically manufactured FPCB designs corresponding to the above hand.}
   
\end{figure*}

We conducted a preliminary user study (N=4) to assess the reliability and comfort of personalized FPCB tactile sensing gloves as compared to a generic, non-personalized glove. Using our pipeline, we generated design components for five right-handed individuals, selecting the hand with median perimeter as the basis for the generic glove. The FPCB layers were manufactured by JLCPCB within nine days of order approval. Following the process in Section \ref{sec:manufacture}, we assembled one generic glove and one personalized glove per participant, resulting in eight gloves total (Fig. \ref{fig:glove-designs}). Manufacturing costs per sensor (two FPCBs) ranged from \$22.78 to \$27.55 (Mean = \$24.93, S.D = \$1.62), with sensor assembly averaging 14 minutes.

Participants performed two experiments using their generic glove ("Glove A") and personalized glove ("Glove B") on their dominant hand. Each experiment was completed with Glove A first, followed by Glove B. In the first experiment, participants wore the glove, pressed their hand onto a table, and removed the glove, repeating this sequence three times while pressure signals were recorded. In the second experiment, participants completed a series of tasks across different stages: opening/closing a book, sliding a mouse (right/left), left-clicking, picking up a water bottle, pouring, straightening the bottle, and setting it down. Each task was repeated 15 times, with video recorded for synchronized task-stage annotation. After the experiments, participants completed a survey assessing obstructiveness, task effort, and frustration for both gloves. The
study was approved by the MIT Committee on the Use of Humans
as Experimental Subjects (COUHES) under the protocol number
2411001482 - Evaluating a Platform for Flexible Printed Circuit Based Tactile Sensing Gloves.

\subsection{Evaluation and Discussion}

\begin{figure*}
    \centering
    \includegraphics[width=1\linewidth]{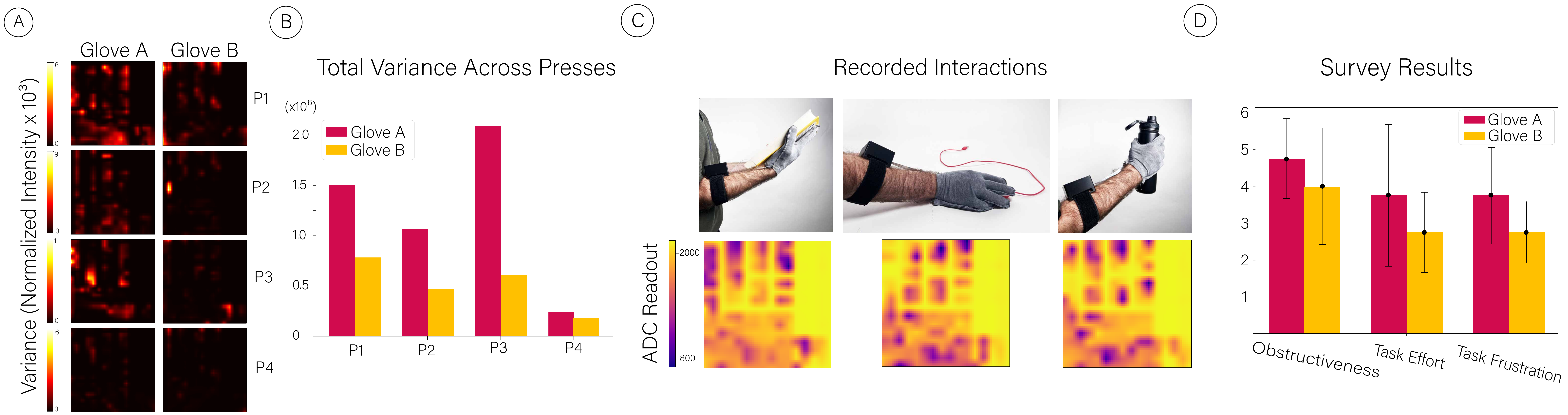}
    \caption{User study results from generic (Glove A) and personalized (Glove B) gloves. (A-B) Personalized gloves demonstrate reduced variance across repeated full-hand presses. (C) Over 42,00 tactile frames are collected while users perform three tasks—opening a book, using a mouse, and pouring water from a bottle. (D) Users rate personalized gloves as less obstructive and report that tasks require less effort and cause less frustration compared to using generic gloves, on average. }
    \label{fig:user-study}
     \Description{Four panels show the results of the two user studies. Panel A: The variance across frames for each glove is displayed as a heatmap for gloves A and B across the four participants. Panel B: Bar plot with participant on the x axis against total variance across presses on the y axis. Side-by-side plots show lower total variance for glove B for each user. Panel C: Paired images and pressure heatmap for three interactions: opening a book, clicking a computer mouse, and holding a water bottle. Panel D: Bar plot depicting survey results, with likert scale rating on the y axis and obstructiveness, task effort, and task frustration on the x axis. Side by side plots for glove A and B show improvements on all three categories, with standard deviations as error bars.}
\end{figure*}

\label{sec:user-eval}

Figure \ref{fig:user-study} presents the results of the two experiments. In the hand press experiment, pressure frames (frame count x 16 x 16) were loaded and the frame with the highest average pressure was identified for each press. The average pressure reading over a 50-frame window (1.3 seconds) centered at this press was calculated, normalized to its minimum and maximum values, converted to pixel intensity (0–255), and interpolated from a 16×16 resolution to 64×64 for enhanced visualization. Figure \ref{fig:user-study}A shows the variance in pixel intensity across all presses for each participant (P1–P4) using both the generic and personalized gloves. Across all participants, the personalized gloves exhibited lower overall pixel variance, suggesting that personalized designs reduce hand placement variability and improve consistency across repeated presses (Fig. \ref{fig:user-study}B). 

We extracted variable-length pressure data sequences for each task stage using start and end times from video annotations, yielding a dataset with 1,076 tactile sequences, totaling over 42,000 frames of pressure data across nine task classes, four participants, and two glove types (Fig. \ref{fig:user-study}C). Excerpts from the dataset demonstrate expected pressure correlations for associated tasks, such as high activation of the index finger region while clicking the mouse and high activation of the palm while holding the water bottle. While analyzing recorded data we observed some loss of sensitivity in the finger regions, averaging 25 taxels per glove due to fractures in the copper traces. We discuss the nature of these fractures further in Section \ref{sec:limitations}.

In our survey, participants evaluated the gloves on a 7-point scale for obstructiveness (1 = Not noticeable, 7 = Very obstructive), task effort (1 = Very low, 7 = Very high), and task frustration (1 = Very low, 7 = Very high). The personalized gloves were rated as less obstructive (Mean = 4.0, SD = 1.58), requiring less effort (Mean = 2.75, SD = 1.09), and causing less frustration (Mean = 2.75, SD = 0.83) (Fig. \ref{fig:user-study}D). While the small participant pool limits generalizability, the results provide interesting insights for future work. For example, participants whose hand sizes most deviated from the generic design showed a stronger preference for personalized gloves and future studies might consider grouping participants by hand size (e.g., small, medium, large), instead comparing personalized designs to generic gloves in multiple sizes. Open-ended feedback also revealed that some participants (P2 and P3) preferred the generic glove due to discomfort from sensor slippage in the personalized design, highlighting the importance of attachment mechanisms for improving comfort and usability. Future research could focus on optimizing these mechanisms.

\begin{figure*}
    \centering
    \includegraphics[width=1\linewidth]{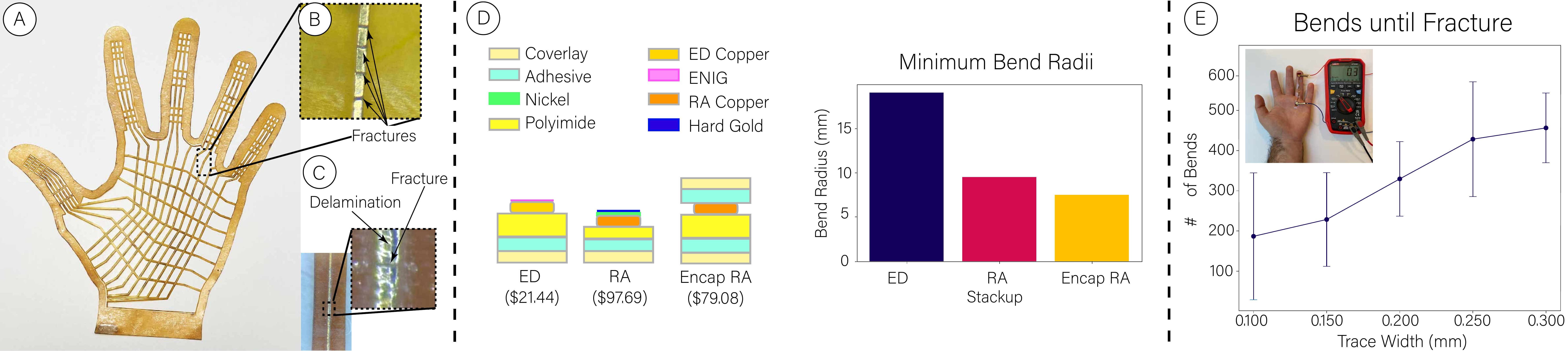}
    \caption{Mechanical investigation. (A) Failure location on the glove. (B-C) Brittle failure was observed in the electrodeposited (ED) copper design while ductile failure and delamination was observed with the rolled annealed (RA) copper design. (D) Comparison of three designs showing the estimated minimum bending radius before failure. In units of $\mu$m, \textbf{ED stackup}: coverlay-12.5, adhesive-15, polyimide-25 , base ED copper-11.6, and an ENIG overlayer-25 nm, from JLCPCB. \textbf{RA stackup}: coverlay-12.5, adhesive-12.5, polyimide-12.5, base RA copper-12, a nickel overlay-3 and hard gold coating-76 nm, from PCBWay. \textbf{Encapsulated RA stackup}: coverlay-12, adhesive-15, polyimide-25, base RA copper-12, adhesive-15, and coverlay-12, from PCBWay. (E) Manual bending tests show increasing trace width improves RA stackup durability, averaging 330 bends to a tight fist before failure.}
    \label{fig:limitations}
    \Description{Three panels showing the results of mechanical investigation. Left: Panel A shows a manufacture fpcb, with a box in the region in between the palm and ring finger highlighting the location of fracture. Panel B shows a microscope image of brittle fracture of the copper trace. Panel C shows a microscope image of the copper trace delaminating before fracturing. Middle: FPCB stackup layers and their material compositions or denoted for electrodeposited, rolled annealed, and encapsulated rolled annealed stack ups. A bar plot of bend radius on the y axis and stack up on the x axis shows the minimum bend radius for each stackup before initial fracture. Right: a line plot depicts results of a manual bending test for copper traces of widths varying from 0.1 mm to 0.3 mm in increments of 0.05, and number of bends ranging from 100 to 600 on the y axis. An image in the top left of the plot shows the experimental setup of a trace taped onto a hand with a multimeter}
\end{figure*}

\section{Limitations and Future Work}

\label{sec:limitations}

Mechanical failure is the greatest limitation of this device, particularly when the glove must bend sharply for difficult or skilled tasks. For this initial study, all FPCBs were manufactured using electrodeposited (ED) copper (JLCPCB) with a coating of electroless nickel and gold (enig), as the lowest-cost option. ED copper is brittle, and we found fractures in the upper palm region after making one or two fists to stress-test the gloves (Fig.~\ref{fig:limitations}A, B). This strongly limits their usability as some gestures cannot be made while wearing them. 

Switching from ED to rolled-annealed (RA) copper with a hard gold coating doubled cost but showed enhanced durability and caused ductile failure (Fig. \ref{fig:limitations}C). Encapsulation of the copper between two coverlayers, as recommended by expert engineers at Epec Engineered Technologies \cite{epec_engineered_technologies}, is predicted to further improve the lifespan of the gloves. A mechanical analysis of bending failure shows that the ED copper design is expected to begin to fail when bent more than a bending radius of 19 mm, about the radius of a ping pong ball, for RA copper 9.5 mm, about the radius of a penny, and for encapsulated RA copper 7.5 mm, about the radius of an AA battery (Fig. \ref{fig:limitations}D). Calculations of bending stress used the standard formula $\sigma=My/I$ for stress $\sigma$, bending moment $M$, maximum height of the copper from the neutral axis $y$, and second moment of area of the cross-section $I$, with tabulated material properties \cite{Yin2016Mechanical, Zhang2011Mechanical}. Finally, increasing trace width was found to more than double the number of cycles the traces can endure for the RA stackup design when directly taped to the hand (Fig. \ref{fig:limitations}E). To further evaluate durability, we conducted a cyclic bending test on a glove fabricated with the RA stackup, where FPCBs were embedded between two silicone layers as proposed in this work. After 10,000 cycles at a 2.5 mm bending radius, no fractures were observed, suggesting that silicone encapsulation may prevent creasing and subsequent trace failure. These design parameters—conductive material, stackup geometry, and trace width—offer tunable trade-offs between cost and durability for future designs.

Finally, while flexible PCBs offer a convenient manufacturing medium, they can be stiffer than purely textile-based approaches. The total thickness of the FPCB glove is approximately 4 mm --- about eight times that of standard nitrile gloves, which may impact wearability. Additionally, like other piezoresistive sensors, we observed motion artifacts caused by bending in the absence of normal force. These artifacts may be mitigated using adaptive strain interference suppression methods, such as those proposed by \citet{Jiang2024}.
\section{Conclusion}

We developed a pipeline for the rapid design of personalized FPCB-based tactile sensing gloves, leveraging accessible board house manufacturing to deliver personalized gloves affordably and quickly. The sensors exhibit sensitivity and consistency comparable to established resistive methods under repeated loads. User evaluations suggest that personalized designs may improve comfort and enable more consistent pressure signals over time. While challenges remain with the mechanical robustness of copper electrodes, stress analysis suggests these can be addressed through tunable design variables, highlighting our methods as a promising approach to making tactile sensing hardware more accessible to a broader research community.
\section{Acknowledgments}

 This work is supported by the MIT-GIST joint program and the Toyota Research Institute.

\bibliographystyle{ACM-Reference-Format}
\bibliography{10_reference}
\end{document}